\documentstyle[eqsecnum,xspace,aps,graphicx]{revtex}

\newcommand{\Mev}{\textrm{MeV}\xspace}
\newcommand{\cerenkov}{Cherenkov\xspace}

\newcommand{\fractional}[1]{{{1 \over #1}}}
\newcommand{\downto}{\mathrel{\raisebox{1.13ex}{$\lfloor$}\!\hbox{$\to$}}}
\newcommand{\gsim}{\mathrel{\rlap{\raisebox{.3ex}{$>$}}
    \raisebox{-.6ex}{$\sim$}}}
\newcommand{\lsim}{\mathrel{\rlap{\raisebox{.3ex}{$<$}}
    \raisebox{-.6ex}{$\sim$}}}

\newcommand{\potassium}{{ {}^{40} \textrm{K}}\xspace}

\newcommand{\thallium}{{{}^{208}_{\phantom{2}81}\textrm{Tl}}\xspace}
\newcommand{\lead}    {{{}^{208}_{\phantom{2}82}\textrm{Pb}}\xspace}

\title{A Search for Lightly Ionizing Particles with the MACRO Detector}

\begin{document}
\draft

\maketitle

\bigskip\begin{center}
  {\bf The MACRO Collaboration}\\
  \nobreak\bigskip\nobreak
  \pretolerance=10000
  M.~Ambrosio$^{12}$, 
  R.~Antolini$^{7}$, 
  G.~Auriemma$^{14,a}$, 
  D.~Bakari$^{2,17}$,
  A.~Baldini$^{13}$, 
  G.~C.~Barbarino$^{12}$, 
  B.~C.~Barish$^{4}$, 
  G.~Battistoni$^{6,b}$, 
  R.~Bellotti$^{1}$, 
  C.~Bemporad$^{13}$
% ...
\\
  \vspace{0.5 cm}
  \footnotesize
  1. Dipartimento di Fisica dell'Universit\`a di Bari and INFN, 70126 
  Bari,  Italy \\
  2. Dipartimento di Fisica dell'Universit\`a di Bologna and INFN, 
  40126 Bologna, Italy \\
  3. Physics Department, Boston University, Boston, MA 02215, 
  USA \\
  4. California Institute of Technology, Pasadena, CA 91125, 
  USA \\
  5. Department of Physics, Drexel University, Philadelphia, 
  PA 19104, USA \\
  6. Laboratori Nazionali di Frascati dell'INFN, 00044 Frascati (Roma), 
  Italy \\
  7. Laboratori Nazionali del Gran Sasso dell'INFN, 67010 Assergi 
  (L'Aquila),  Italy \\
  8. Depts. of Physics and of Astronomy, Indiana University, 
  Bloomington, IN 47405, USA \\
  9. Dipartimento di Fisica dell'Universit\`a dell'Aquila  and INFN, 
  67100 L'Aquila,  Italy \\
  10. Dipartimento di Fisica dell'Universit\`a di Lecce and INFN, 
  73100 Lecce,  Italy \\
  11. Department of Physics, University of Michigan, Ann Arbor, 
  MI 48109, USA \\      
  12. Dipartimento di Fisica dell'Universit\`a di Napoli and INFN, 
  80125 Napoli,  Italy \\       
  13. Dipartimento di Fisica dell'Universit\`a di Pisa and INFN, 
  56010 Pisa,  Italy \\ 
  14. Dipartimento di Fisica dell'Universit\`a di Roma ``La Sapienza" and INFN, 
  00185 Roma,   Italy \\        
  15. Physics Department, Texas A\&M University, College Station, 
  TX 77843, USA \\      
  16. Dipartimento di Fisica Sperimentale dell'Universit\`a di Torino and INFN,
  10125 Torino,  Italy \\       
  17. L.P.T.P., Faculty of Sciences, University Mohamed I, B.P. 524 Oujda, Morocco \\
  18. Dipartimento di Fisica dell'Universit\`a di Roma Tre and INFN Sezione Roma Tre, 
  00146 Roma,   Italy \\        
  $a$ Also Universit\`a della Basilicata, 85100 Potenza,  Italy \\
  $b$ Also INFN Milano, 20133 Milano, Italy\\
  $c$ Also Istituto TESRE/CNR, 40129 Bologna, Italy \\
  $d$ Also Scuola Normale Superiore di Pisa, 56010 Pisa, Italy\\
  $e$ Also Universit\`a di Trieste and INFN, 34100 Trieste, 
  Italy \\
  $f$ Also Dipartimento di Energetica, Universit\`a di Roma, 
  00185 Roma,  Italy \\
  $g$ Also Institute for Nuclear Research, Russian Academy
  of Science, 117312 Moscow, Russia \\
  $h$ Also Institute for Space Sciences, 76900 Bucharest, Romania \\
  $i$ The Colorado College, Colorado Springs, CO 80903, USA\\
  $l$ Also INFN Catania, 95129 Catania, Italy\\
  $o$ Also Department of Physics, Pennsylvania State University, University Park,
  PA 16801, USA \\
  $p$ Also Department of Physics, James Madison University, Harrisonburg,VA 22807,
  USA\\
  $q$ Also Dipartimento di Fisica dell'Universit\`a della Calabria,  Rende (Cosenza),
  Italy\\       

\end{center}

\begin{abstract}
  
  A search for lightly ionizing particles has been performed with the
  MACRO detector.  This search was sensitive to particles with charges
  between $\fractional{5}$$e$ and close to the charge of an electron,
  with $\beta$ between approximately 0.25 and 1.0. Unlike previous
  searches both single track events and tracks buried within high
  multiplicity muon showers were examined.
  
  In a period of approximately one year no candidates were observed.
  Assuming an isotropic flux, for the single track sample this
  corresponds to a 90\% C.L. upper flux limit $\Phi \leq 9.2 \times
  10^{-15} \; cm^{-2} sec^{-1} sr^{-1}$ .
  
\end{abstract}

\newpage

\section{Introduction}

Ever since Robert Millikan's historic experiment determined that the
charge on matter comes in discrete units~\cite{Millikan},
experimenters have spent much time and effort first determining the
precise value of that charge, and later trying to observe instances in
nature where anything other than an integer multiple version of that
charge exists.

The first hint that such objects might be present in nature were the
results obtained from the deep inelastic scattering experiments at
SLAC during the late 1960's~\cite{DIS}.  These experiments first
demonstrated that nucleons do in fact have sub-structure.  By
exploring the structure functions in these scattering experiments, it
was discovered that protons and neutrons were constructed of smaller
point-like partons, and that there were three charge-bearing partons
in each of the proton and the neutron~\cite{partons}.

This observed parton structure fit well into the quark model
previously proposed by Gell-Man and
Zweig~\cite{quark-model,zweig401,zweig412}.  Although in this model
the quarks which make up the baryons and mesons have fractional
charge, they are always combined in a way that results in an
integrally charged baryon or meson.

Despite decades of searching no one has yet observed a quark free of
its ever-present neighbors.  Also, the search for electrons or other
leptonic type particles with fractional charge has been in vain.
These include larger and more sophisticated versions of Millikan's oil
drop experiment, searches in bulk matter, experiments at accelerators,
and searches in the cosmic
radiation~\cite{Jackson:1989bw,Lyons:1985,Klapdor,Jones,Halyo:1999wq}.
A clear observation of fractional charge would be extremely important
since, depending on the type of particle seen, it might mean that
confinement breaks down under some circumstances or that entirely new
classes of particles exist.

In Grand Unified Theories it is relatively easy to accommodate
fractional charge in color singlets by extending the unification group
from \textbf{SU(5)} to a larger group.  For example, an extension to
\textbf{SU(7)} allows for charges of $\fractional{3}$~\cite{Frampton}
,another which allows $\fractional{3}$~e charge leptons has gauge group
\textbf{SU(5) $\times$ SU(5)'}~\cite{Barr}.  Other Grand Unified Theory groups
have been considered which allow for fractional charge, including
\textbf{SU(8)}~\cite{Yu:1984}, \textbf{SO(14)}~\cite{Yamamoto:1983},
and \textbf{SO(18)}~\cite{Dong:1983}.  Further, some theories of
spontaneously broken QCD have also predicted free
quarks~\cite{DeRujula:1978}, although these quarks would probably be
contained in super-heavy quark-nucleus complexes with large
non-integral charge.

This paper presents the results of a search for penetrating, weakly
interacting particles with fractional charge in the cosmic radiation
with the MACRO detector.  A more detailed description of this analysis
can be found in~\cite{walter-thesis}.  Since a particle of charge Q
has a rate of energy loss by atomic excitation and ionization
proportional to $Q^2$, particles of a given velocity with fractional
charge deposit less energy in a detector than particles with unit
charge.  So, for example, a particle traveling at relativistic speed
with charge of $\fractional{3}$~e will have an energy deposition only
$\fractional{9}$ that of the muon.  For this reason we call such
particles \textit{lightly ionizing particles}(LIPs).  A quark of the
standard model also interacts via the strong force and would not be
able to penetrate large amounts of material; thus this search is only
sensitive to penetrating lightly ionizing particles.

\section{Experimental Setup}

The MACRO detector is a large($\approx$~10000~$m^2$~sr) underground
scintillator and streamer tube detector and has been described in
detail elsewhere~\cite{technical,streamer-mono}.  Due to MACRO's large
size, fine granularity, high efficiency scintillator, and high
resolution tracking system, it is uniquely suited to look for LIPs.
In order to take advantage of this situation a special LIP trigger
system has been built.

Using the lowest level energy-based scintillator trigger available in
MACRO, it allows a search for particles which interact
electro-magnetically but deposit much smaller amounts of energy in the
scintillator counters than minimum ionizing muons.  The inputs are the
individual counter low energy triggers produced in the PHRASE (one of
the gravitational collapse triggers), which have a trigger threshold
of about 1.2~\Mev.  Since a typical muon energy loss is about 40 MeV,
this trigger threshold allows a search for particles losing less than
1/25 of this.

Streamer tubes are more efficient at triggering on LIPs than the
scintillator system.  The key to the good sensitivity of the streamer
tubes, even to extremely small amounts of ionization, is that even a
single ion-electron pair produces a full streamer with reasonable
probability.

The measured single ion-pair efficiency for the MACRO tubes, gas
mixture, and operating voltage is over 30\%, which is consistent with
earlier work~\cite{streamer-single}. Since selected tracks are
required to cross at least 10 streamer tube planes and a LIP trigger
only requires that any 6 of them fire, the streamer tube triggering
probability is over 99\% for the range of charges considered in this
search.

The LIP trigger uses field programmable gate array circuits to form
coincidences between counters in the three horizontal planes of MACRO
scintillator.  The resulting accidental coincidence rate of
approximately 10 Hz would overload the data acquisition and storage
system and so it is reduced by requiring a coincidence with at least 6
streamer tube planes in the bottom part of the detector.

Since a well-reconstructed streamer tube track is required in the
final off-line analysis, the streamer tube trigger requirement does
not reduce the efficiency of the search, although it reduces
accidental coincidences to an acceptable level. The LIP trigger stops
the 200 MHz waveform digitizer(WFD) system and causes the data
acquisition system to readout the waveforms of all the counters
involved in the trigger.

The use of this trigger allows a physics search for LIPs which is
unique in many ways.  Some of the main features which distinguish it
are as follows:

\begin{enumerate}

\item Sensitivity down to $\fractional{5}$ equivalent fractional charge.
  Previous experiments have only checked for particles with charge 
  $\gsim \fractional{3}$~\cite{Lyons:1985}.
  
\item Good acceptance from $\beta$~=~0.25 - 1.0 . Particles which have
  a velocity lower than 0.25c are not guaranteed to pass through the
  detector quickly enough to insure that the LIP trigger will detect a
  coincidence in the faces of the scintillator system.  The lowest
  flux limits for LIPs now come from the very large water \cerenkov
  detector in Japan (Kamiokande)\cite{Mori:1991}.  However, because of
  the nature of the \cerenkov process, water detectors are only
  sensitive to particles with $\beta \gsim 0.8$.
  
\item Size of detector.  The MACRO detector presents
  $\approx$~800~$m^2$ of fiducial area to downward-going particles.
  The \cerenkov search at Kamiokande presents a nominal detection area
  of 130~$m^2$\cite{Mori:1991}. The best results from
  scintillator-based experiments come from even smaller detectors.
  The search by Kawagoe \textit{et al.}~\cite{Kawagoe:1985} relied on
  a detector of only 6.25 $m^2$.
  
\item The possibility of searching within large multiple muon bundles
  for fractional charge.  Because of the size and granularity of the
  MACRO experiment, it is possible to isolate tracks located in muon
  bundles containing on the order of 20 muons, and to check their
  energy deposition to see whether they are consistent with LIPs.  For
  both smaller experiments and non-granulated experiments (such as
  single large volume water experiments like Kamiokande), multiple
  muon events are rejected from the data sample.  If fractionally
  charged particles were being produced in high energy collisions in
  the upper atmosphere, previous experiments may have missed the
  signature due to the particles being buried in the high-multiplicity
  shower.
  
\item Use of high resolution waveform digitizers for energy and timing
  reconstructions.  At a trigger threshold of $\approx$ 1.2~\Mev each
  scintillator counter fires at approximately 2~kHz.  The traditional
  ADC/TDC system is susceptible to errors associated with false
  starts at this rate (see for example~\cite{Napolitano:1982}). A
  small pulse triggering the ADC/TDC system just prior to a large
  pulse can result in partial integration of the large pulse,
  producing a fake low ionization event.
  
\item Use of a high precision limited streamer tube tracking system.
  Previous underground
  experiments~\cite{Mori:1991,Kawagoe:1985,Aglietta:1994} did not have
  independent tracking systems. Since muons that clip the corners of
  scintillating volumes can be an important source of background, the
  use of a tracking system is essential for the performance of a low
  background search.  In addition, without a tracking system it is
  hard to recognize the cases where the actual tracks pass between
  volumes and accompanying soft gamma rays enter into the
  scintillating volumes.  This can be a source of
  background~\cite{Aglietta:1994}.  The use of a tracking system is
  also one of the reasons that MACRO can look for fractional charge in
  high multiplicity muon bundles.

\end{enumerate}

\section{Data Analysis}

The data for this search comes from two periods.  The first ran from
July 24th to October 12th of 1995, and the second from December
17th 1995 to November 16th 1996.  These were both periods of
uninterrupted waveform and LIP operation with the entire MACRO
detector.  The live-time varied for sub-sections of the detector and
the longest live-time was 250 days.

\subsection{Low Energy Reconstruction}
\label{sec:low-energy-calibration}

Triggering at very low thresholds is challenging.  While previous
searches have restricted themselves to $\fractional{3}e$, this search
reaches $\fractional{5}e$.  For particles with average path lengths
through MACRO scintillator counters the energy deposited is about

\begin{equation}
  40\ \Mev \times \left( \fractional{5} \right)^2 \approx 1.6\ \Mev .
\end{equation}

\noindent
Therefore, in order to be able to reconstruct LIPs which pass through
MACRO, it is necessary to reconstruct energies between 1.5 and 40
\Mev. 

\begin{figure}[p]
  \begin{center}
    \leavevmode
    \includegraphics[width=5.0in]{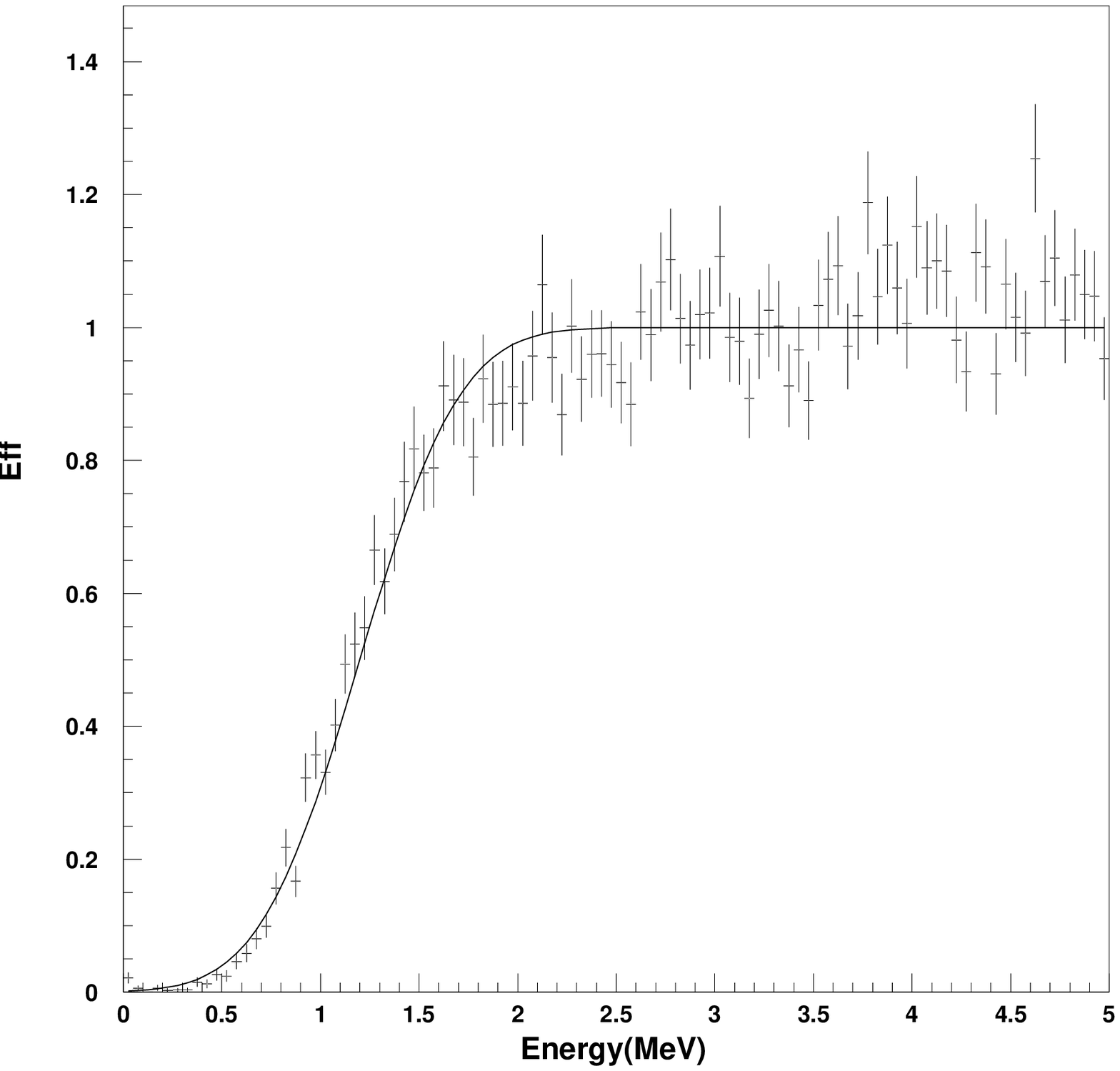}
    \caption{The measured efficiency of triggering the low energy PHRASE
      trigger and the LIP trigger as a function of energy. Some
      measured efficiencies are greater than 100\% because the
      normalization factor used is only an estimate of the true
      normalization as a function of energy.
}
    \label{fig:eff-dist}
  \end{center}
\end{figure}

The triggering threshold of the LIP trigger was measured by using
muons which passed through small amounts of scintillator in the MACRO
detector, and thus deposited small amounts of energy.  The measured
triggering efficiency is shown in figure~\ref{fig:eff-dist}; it is
100\% above $\approx$~2~\Mev, 50\% above 1.2~\Mev.

Each scintillator counter used in the analysis was calibrated using
naturally occurring low-energy $\gamma$-rays.  The most important of
these $\gamma$-rays for the calibration is the 2.6~\Mev line from the
radioactive decay-chain:

\begin{eqnarray}
  \label{eq:thallium}
  \thallium &           &   \nonumber \\
  \downto   & \lead^{*} & + \ \beta^{-} + \overline{\nu_e} \nonumber \\
            & \downto   &   \lead     + \gamma\ (2.6 \Mev) .
\end{eqnarray}

\begin{figure}[!hbtp]
  \begin{center}
    \leavevmode
    \includegraphics[width=5.0in]{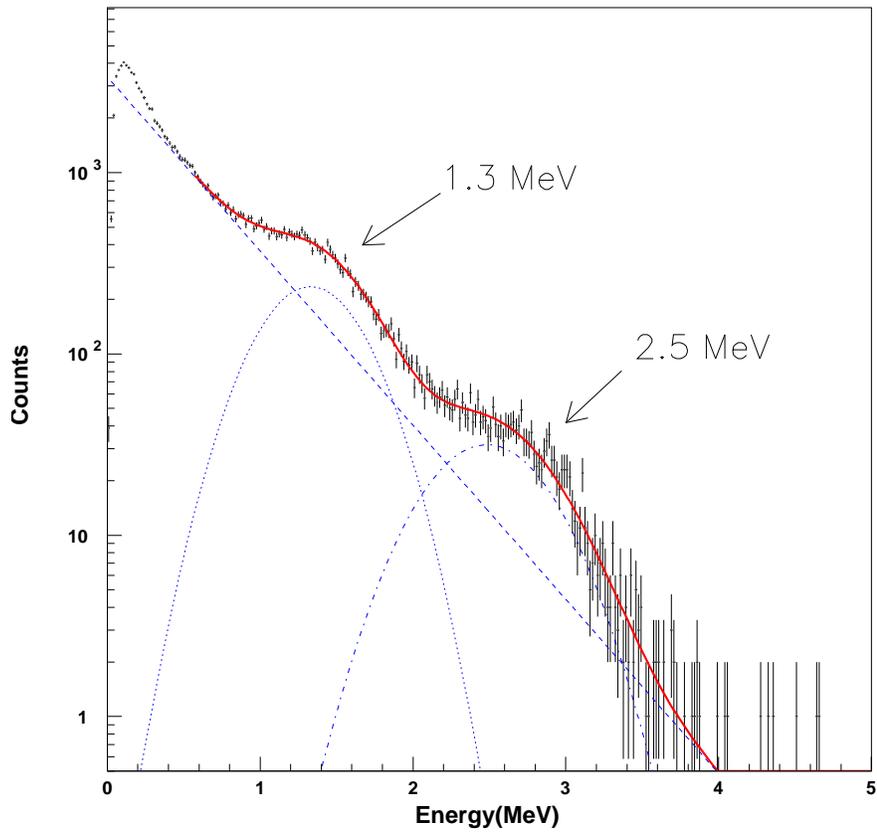}
    \caption{A fit to low energy WFD data with a 
      falling radioactive spectrum, and a Gaussian associated with
      both the 2.6 \Mev $\gamma$(Tl) and 1.4 \Mev $\gamma$(K) line.
      Each energy bin is 16.7 keV wide.  The eight parameters of the
      fit are the normalization and slope for an exponential and the
      normalization, mean, and width for the two Gaussians. }
    \label{fig:scintillator-spectrum}
  \end{center}
\end{figure}

After every event which causes a readout of the WFDs, one millisecond
worth of WFD data is collected for every counter involved in the
event.  For fast particles such as muons only the first few
microseconds of the WFD data is relevant.  The rest of the data is
recorded in order to search for slowly moving particles such as
magnetic monopoles.  The one millisecond of data contains small pulses
caused by naturally occurring radioactivity.  By looking at these
radioactivity pulses we can reconstruct the low energy spectrum.
Figure~\ref{fig:scintillator-spectrum} shows this spectrum for one of
the MACRO scintillator counters.

The solid line is a fit to a falling radioactivity spectrum plus two
gaussians, one associated with the 2.6~\Mev $\thallium$ line, and the
other, with the 1.4~\Mev $\potassium$ line.  A full GEANT Monte Carlo
was performed to determine where the absolute energies of the lines in
this spectrum should be, and the information from the fit is used to
make a calibration constant to convert between observed PMT signal
measured in the waveforms and deposited energy.

Since one to five \Mev is the important signal region for the LIP
search, reconstructing the low energy spectrum in this region is proof
that we can also reconstruct LIPs in this region.  For this reason, we
require a counter to have a good calibration in order to use it for
the LIP analysis.  Aside from a very few non-functional scintillator
counters, in practice, what this means is that only the counters
placed in three horizontal planes were used, and the counters in
vertical planes were not.

\subsection{Time Reconstruction }

It is important to determine an event's longitudinal position in a
counter from its WFD data. Calibration events as described in
section~\ref{sec:low-energy-calibration} have no associated streamer
tube track, and so this is the only source of the information
necessary to correct for the light attenuation of the scintillator.
For particles passing through the detector, we require consistency
between the longitudinal position of the event independently
determined by the streamer tubes and the PMT signals. This reduces
the background due to accidental coincidences between a small
radioactivity pulse somewhere in the counter followed by a muon
passing through a crack in the detector.  The width of the position
resolution determines how tightly this cut can be made.

\begin{figure}[p]
  \begin{center}
    \leavevmode
    \includegraphics[width=5.00in]{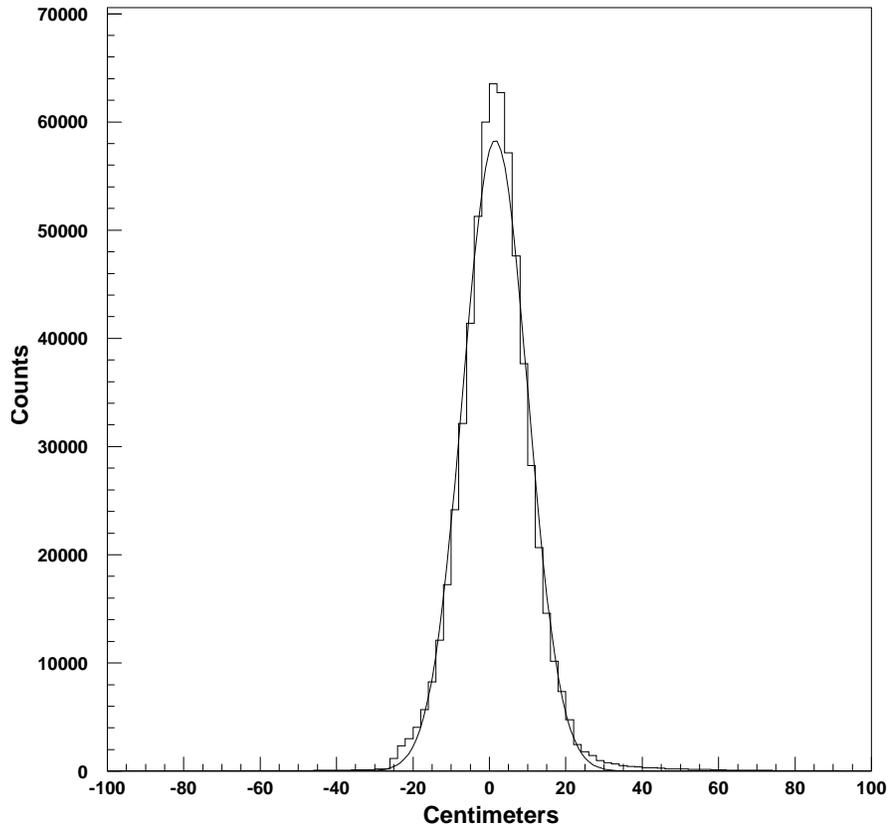}
    \caption{Difference in position calculated by the streamer tubes
      and that by the PMT signals for a sample of the muon data. The
      r.m.s deviation from the mean (sigma) is 8.5 cm. All counters
      used in the analysis are included in this histogram; individual
      counters have smaller sigmas.}
    \label{fig:netDiffCor}
  \end{center}
\end{figure}

The longitudinal position in a counter of an event can be calculated
using the WFD information with the expression:

\begin{equation}
  \label{eq:position}
  pos = {\Delta t \times v \over 2 } ,
\end{equation}

\noindent
where $\Delta t$ is the difference in time between the pulses on the
two sides of the counter (as measured by the waveforms), and $v$ is
the effective speed of light in the counter.
Figure~\ref{fig:netDiffCor} shows the difference between the positions
of muons passing through a scintillator counter calculated by the
streamer tube tracking system and that calculated by the WFD system
for all of the scintillator counters used in the analysis.

These timing results were obtained by first performing a software
simulation of a constant fraction trigger~\cite{Leo} to obtain an
initial estimate of the longitudinal position.  This circuit triggers
at the point on the leading edge of a pulse which is a fixed fraction
of the maximum height of the pulse.  In order to estimate at what time
the pulse crosses the fixed fraction of the maximum peak voltage (20\%
was used for this analysis) a simple linear fit was used between the
two samples closest to the point of crossing.

A neural network was then used to further refine the estimate of the
longitudinal position.  The neural network was trained with a sample
of events using the position obtained from the streamer tubes.  We
chose to use a neural network since we did not find an alternative
which provided the same or better precision and was less
computationally intensive.  A more detailed description of the network
used can be found in~\cite{walter-thesis}.

\section{Search Results}

After calibration, the data set was examined for LIPs in both single
and multiple track events.  In order to be considered in the analysis,
an event had to satisfy three requirements: the LIP trigger must have
fired; at least one track must have been reconstructed in the streamer
tube system; and finally, at least one of the reconstructed tracks
must have passed through counters in the top, center, and bottom of
the detector.  There were approximately 1.3~million events which
satisfied these requirements.  The data set was broken into two
exclusive pieces, a single track and a multiple track set, with
approximately 90\% of the events being in the single track sample.

\begin{figure}[p]
  \begin{center}
    \leavevmode
    \includegraphics[width=5.0in]{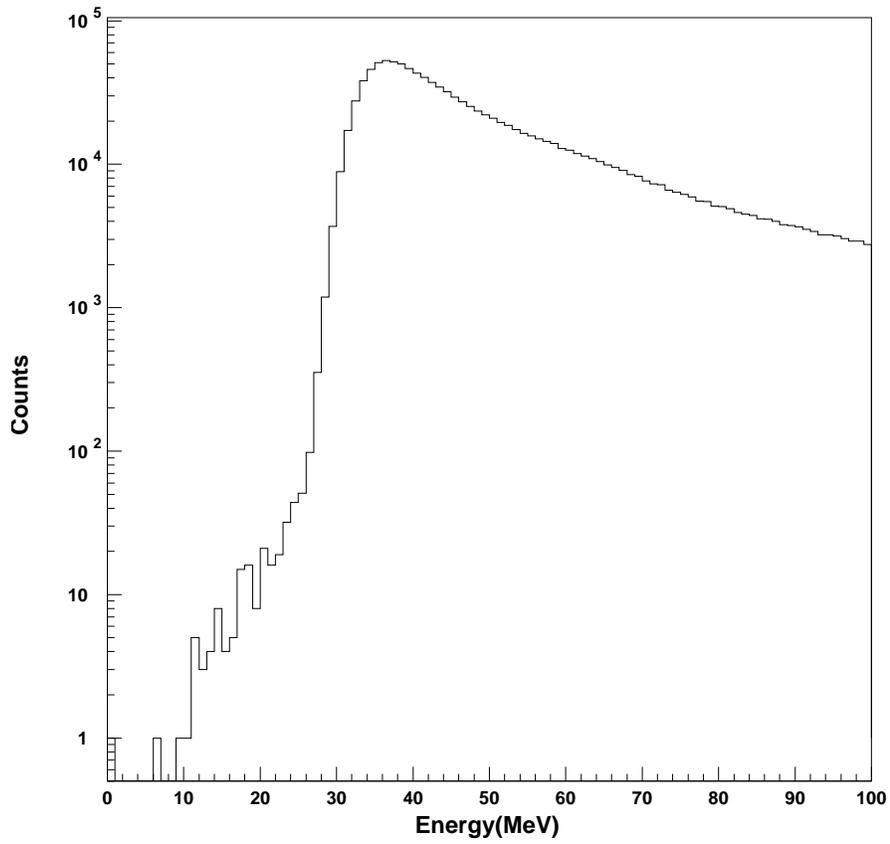}
    \caption{The maximum energy reconstructed in any counter on the
      track. Only an event in which every counter has a low energy will
      show up as having low energy in this histogram.  The
      reconstructed energy has been scaled to a 19~cm. path length for
      all events.}
    \label{fig:maxEnergy}
  \end{center}
\end{figure}

Each of the selected events was then examined to determine its rate of
energy loss in the scintillator.  For each of the counters that a
selected particle passed through, the reconstructed energy was scaled
to a common path length of 19 cm, the distance a vertical muon passing
through a scintillator counter traveled. To reduce the chance that
anomalies would affect the result, the maximum energy in any of the
counters was used as a measure of the particle's energy loss.
Figure~\ref{fig:maxEnergy} is a histogram of this distribution for all
of the tracks(in both the single and multiple track sample) that passed
the selection criteria.

The trigger becomes more than 60\%efficient at about 1.2~\Mev and
quickly rises to 100\% efficiency. Then, at about 20~\Mev, the
efficiency of this search quickly drops to zero because a cut must be
made to reject muons.  Before any cuts, there are events in the region
where LIPs would be expected to appear ($\lsim$ 20~\Mev).  These
result from two classes of reconstruction errors.  First, there are
cases where tracks passed close to the edge of a scintillator counter
or very close to a phototube and the energy was incorrectly
reconstructed.  We therefore also exclude tracks which at their center
in the scintillator volume are located in the final 10~cm of a
scintillator counter. By requiring that all tanks hit by the track
have this fiducial requirement, the number of events in the single
track sample is reduced by $\approx$ 4\%.

Secondly, there are events in which the position reconstructed by
timing in the scintillator counter is inconsistent with that obtained
by the streamer tube tracking system, possibly due to random noise in
the streamer tubes confusing the tracking algorithm.  We require that
the position of particle passage as reconstructed in the streamer
tubes agrees with the position as reconstructed by the neural network
timing procedure to within $\pm$~45 cm, which is about 3~$\sigma$ for
energy depositions smaller than 5~\Mev.  This cut removes 1.8\% of the
data.

\begin{figure}[p]
  \begin{center}
    \leavevmode \includegraphics[width=5.0in]{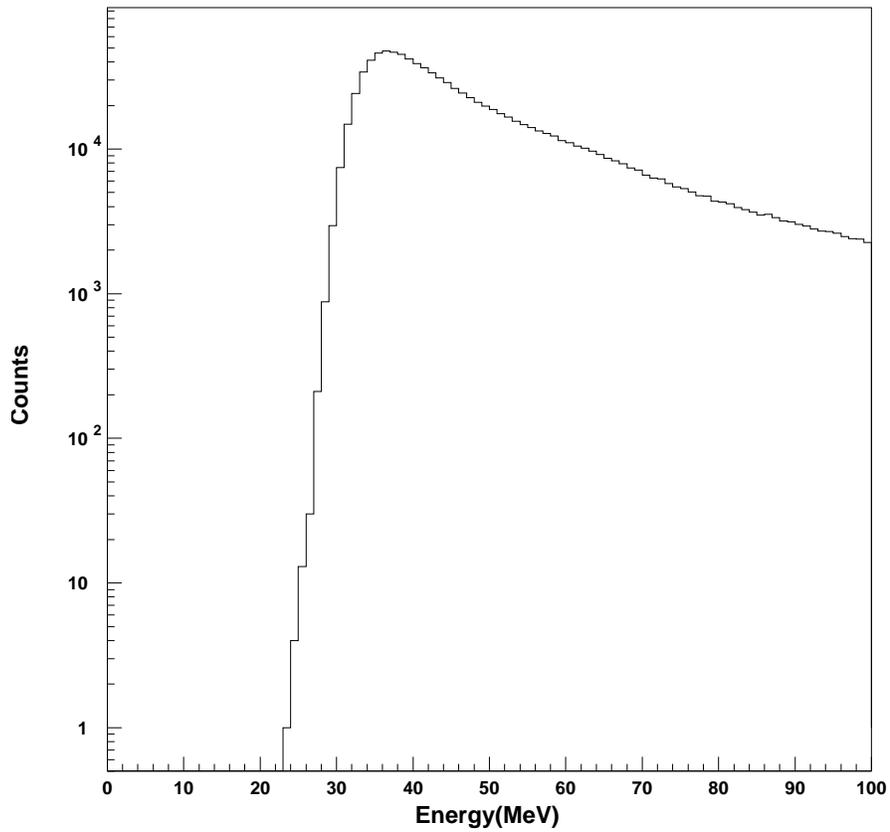}
    \caption{The maximum energy reconstructed in any counter 
      on the track for events in the single track sample. The streamer
      tube and scintillator position reconstruction have been required
      to agree to within $\pm$ 45 cm. and fiducial cuts in the
      scintillator volume have been applied.}
    \label{fig:maxEnergy-fiducial}
  \end{center}
\end{figure}

Figure~\ref{fig:maxEnergy-fiducial} is the distribution of the maximum
counter energy on a track for all of the single muon tracks considered in
the analysis after the fiducial and position agreement cut.  The
expected signal region for LIPs is below 20~\Mev.

\begin{figure}[p]
  \begin{center}
    \leavevmode
    \includegraphics[width=5.0in]{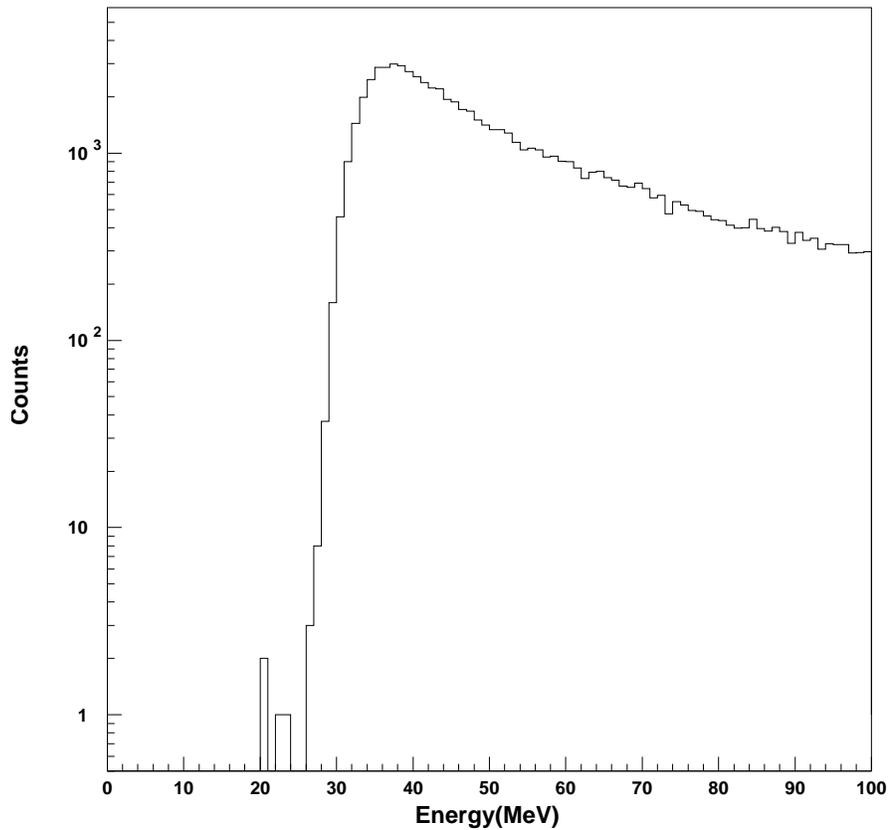}
    \caption{The maximum energy reconstructed in any counter
      on the track of the event for the multiple track sample. The
      streamer tube and scintillator position reconstruction have been
      required to agree to within $\pm$ 45 cm, and fiducial cuts in
      the scintillator volume have been applied. The events with the
      three lowest energies arose from falsely reconstructed tracks in
      the streamer tube system.  There are no real tracks associated
      with these events.}
    \label{fig:maxEnergy-multiple}
  \end{center}
\end{figure}

Figure~\ref{fig:maxEnergy-multiple} is the same distribution for the
multiple track sample.  There are four events in the multiple tracks
sample with maximum deposited energies between 20 and 23~\Mev.  The
minimum entry in the distribution for the single track sample is
23~\Mev.  These four events were examined by hand.  All four were
reconstructed as double muons by the tracking algorithm.  In three
cases, the tracking algorithm failed and assigned a track where one
really did not exist.  This nonexistent track intersected counters
that were actually hit, but the calculated path lengths with the fake
track were incorrect.  The fourth event had a maximum energy loss of
23~\Mev.  This event shows no anomalies and is consistent with the
lowest energy seen in the single track sample.

\section{Conclusions}

In the approximately one year of running that this search covers, no
candidates for LIPs were observed.  This search was sensitive to
particles with charges greater than $\fractional{5}$$e$ and $\beta$ 
between approximately 0.25 and 1.0. Unlike previous experiments, this
search attempted to find LIPs in both single track events and buried
among the tracks of multiple muon showers.

For the single track sample, the assumption of an isotropic flux
yields a 90\% C.L. upper flux limit of $\Phi \leq 9.2 \times 10^{-15}
\; cm^{-2} sec^{-1} sr^{-1}$.

Once again, it should be emphasized that the energy loss considered
for particles in this search is due solely to atomic excitation and
ionization. If LIPs are present in the cosmic rays and they interact
strongly as well as electro-magnetically they will not be able to
travel through enough rock to reach the MACRO detector before they
interact strongly.  Only if strongly interacting LIPs were produced in
the rock very near the detector would this search be sensitive to
them.

The two best experiments to compare this result with are the LSD
experiment~\cite{Aglietta:1994} and the Kamiokande
experiment~\cite{Mori:1991}.  While LSD had the best
scintillator-based limit in the world prior to this experiment,
Kamiokande has the lowest limit.  Both of these experiments only claim
sensitivity to $\fractional{3}$~$e$ and $\frac{2}{3}e$ charged
particles.

Table~\ref{tab:limits} summarizes the limits of this search in
comparison to other searches.  For the entries marked ``Not Quoted'',
the experiments do not report a limit for that charge although the
experiment should have been sensitive to that energy deposition.  At
least in the case of LSD there were two candidates in the
$\fractional{2}e$ region which were ignored because they were not
considering $\fractional{2}e$ charged particles.  In the Kamiokande
experiment only $\fractional{3}e$ and $\frac{2}{3}e$ were searched
for.

\begin{figure}[p]
  \begin{center}
    \leavevmode
    \includegraphics[width=5.0in]{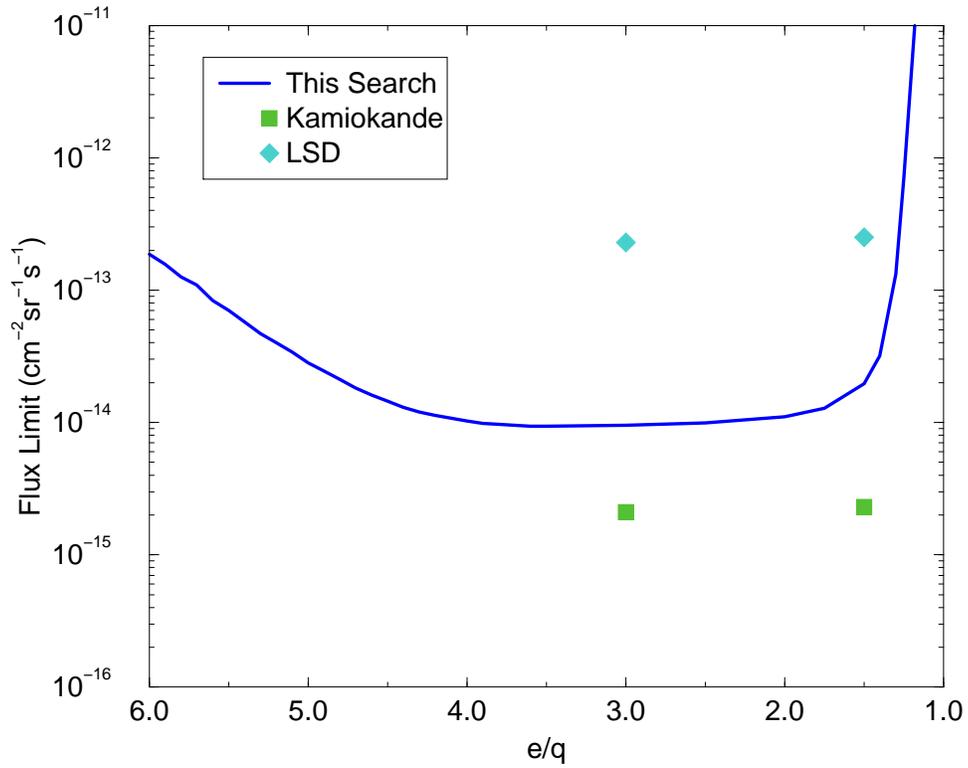}
    \caption{The upper limit on LIP fluxes at 90\% confidence level
      established by this search.  q is the charge of the LIP.  Also
      shown are the limits from the searches done at the Kamiokande
      and LSD experiment.  Unlike those experiments we report a limit
      for a continuous range of charges.  For this analysis the stated
      flux limits are valid for a beta range of 0.25-1.0 .}
    \label{fig:limit}
  \end{center}
\end{figure}

Unlike the other two searches this search is sensitive to a continuous
range of charges from $\fractional{5}e$ to close to the charge of an
electron.  This limit is shown in figure~\ref{fig:limit}.  This search
had no candidates and required hand scanning of only 3 in 1.2 million
events.

In order to compare flux limits for LIPs from different experiments
one must keep several factors in mind.  First of all, this is a
limit on the flux of local LIPs at the site of the detector.
Different mechanisms for LIP production result in different properties
for their flux.  One possibility is that the LIP particles are
produced very close to the detector by some unknown neutral particle
or mechanism. In this case, one could indeed expect a location
independent, isotropic flux.

However, for the more general case of LIP production far away from the
detector, one expects different fluxes of LIP particles in different
underground locations.  At each detector site there will be a unique
and non-trivial angular distribution, because of different rock
thickness above the detectors.  This will be true if the LIP particles
are produced near the detector in high energy muon showers, in cosmic
ray showers in the atmosphere, or if they are impinging on the Earth
from outer space.  

Note that only particles above some minimum energy can reach an
underground detector from the atmosphere, because of the ionization
loss in the Earth.  For the case of MACRO, which has a minimum depth
of 3300 meters of water equivalent, the initial energy of a
$\fractional{5}$$e$ charged particle before it enters the earth must
be $\ge 20$~GeV.  In comparison, the Kamiokande experiment has an
overburden of 2700~meters of water equivalent, and the LSD experiment
is covered with 5000~meters of water equivalent so the energy
thresholds should be correspondingly lower and higher respectively.

In a general discussion such as this one we can only make some
qualitative remarks.  If the LIP particles are produced in the
atmosphere they should not arrive from directions below the horizon.
A $\fractional{5}$$e$ charged particle would travel 25 times as far as a
muon by virtue of its reduced energy loss, but that distance is still
very small compared to the diameter of the earth.

To compare the results of the different experiments one should
therefore, in principle, consider a particular physical model of
production of the particles, a detailed description of the material
above the detectors, and the detector acceptances (including their
angular dependences).

\acknowledgements

We gratefully acknowledge the support of the director and of the staff
of the Laboratori Nazionali del Gran Sasso and the invaluable
assistance of the technical staff of the Institutions participating in
the experiment. We thank the Istituto Nazionale di Fisica Nucleare
(INFN), the U.S. Department of Energy and the U.S. National Science
Foundation for their generous support of the MACRO experiment. We
thank INFN, ICTP (Trieste) and NATO for providing fellowships and
grants (FAI) for non Italian citizens.



\newpage

\newpage
\begin{table}[p]
  \begin{center}
    \leavevmode
    \begin{tabular}{|c|c|c|c|c|c|}  \hline
              \multicolumn{6}{|c|}{Charge} \\ \cline{2-6}
      Search &$\fractional{5}$&$\fractional{4}$&$\fractional{3}$& $\fractional{2}$ & $\frac{2}{3}$ \\ \hline
      This Search & $2.8 \times 10^{-14}$ &  $1.0 \times 10^{-14}$ & $ 9.5 \times 10^{-15}$&  $1.1 \times 10^{-14} $& $2.0 \times 10^{-14}$ \\
      LSD         & - & - & $2.3\times 10^{-13}$ &Not Quoted& $2.7\times10^{-13}$ \\
      Kamiokande & - & - & $2.1\times 10^{-15}$ &Not Quoted&  $2.3\times10^{-15}$ \\  \hline
    \end{tabular}
    \caption{A summary of limits in LIP searches expressed in units of 
      $cm^{-2} sec^{-1} sr^{-1}$.  This limit(MACRO) is compared with
      limits from the water \cerenkov \ Kamiokande experiment and the
      scintillator based LSD experiment.  The MACRO experiment is
      alone in setting a limit on $\frac{1}{5}e$ and $\frac{1}{4}e$
      charged particles. A ``-'' in the table means that the listed
      experiment was not sensitive to the relevant charge while ``Not
      Quoted'' means that while in principle the detector was
      sensitive, the authors chose not to report a limit for that
      charge.}
    \label{tab:limits}
  \end{center}
\end{table}

\end{document}